# The imperative for reproducibility in building performance simulation research

Christian Ghiaus *

INSA Lyon, CNRS, CETHIL UMR5008, 69621 Villeurbanne, France

## Abstract

Building Performance Simulation (BPS) uses advanced computational and data science methods. Reproducibility, the ability to obtain the same results by using the same data and methods, is essential in BPS research to ensure the reliability and validity of scientific results. The benefits of reproducible research include enhanced scientific integrity, faster scientific advancements, and valuable educational resources. Despite its importance, reproducibility in BPS is often overlooked due to technical complexities, insufficient documentation, and cultural barriers such as the lack of incentives for sharing code and data. This paper encourages the reproducibility of articles on computational science and proposes to recognize reproductible code and data, with persistent Digital Object Identifier (DOI), as peer-reviewed archival publications. Practical workflows for achieving reproducibility in BPS are presented for the use of MATLAB and Python.

## 1 Introduction

### 1.1 Place of building performance simulation (BPS) in scientific paradigms

Oversimplifying, scientific research has evolved through four primary paradigms: empirical, theoretical, computational, and data science. Empirical science, grounded in observation and experimentation, forms the basis of our initial understanding of natural phenomena. Theoretical science employs mathematical models and abstractions to explain these phenomena. Computational science uses computers to solve complex scientific problems that are otherwise intractable. Data science integrates empirical data with advanced computational methods to uncover patterns and insights (Hey, Tansley, & Tolle, 2009).

Building Performance Simulation (BPS) leverages computational science and, increasingly, data science. It uses sophisticated software tools to model and analyze various aspects of building performance, including energy efficiency, thermal comfort, and indoor air quality.

---

* CONTACT Christian Ghiaus  christian.ghiaus@insa-lyon.fr

Table 1 *Number and percentage of some software mentioned in 570 papers published by Journal of Building Performance Simulation between 2008 and 2024 (search results on https://www.tandfonline.com).*

| Software | Mentioned: anywhere in the paper | | in the abstract | |
|---|---|---|---|---|
| | Number | Percentage | Number | Percentage |
| EnergyPlus | 265 | 46.49 % | 62 | 10.88 % |
| TRNSYS | 143 | 25.09 % | 23 | 4.04 % |
| ESP-r | 81 | 14.21 % | 11 | 1.93 % |
| EDA-ICE | 29 | 5.09 % | 0 | 0 |
| Comfie | 5 | 0.88 % | 1 | 0.18 % |
| Modelica | 65 | 11.40 % | 29 | 5.08 % |
| COMSOL | 12 | 2.11 % | 1 | 0.18 % |
| MATLAB | 127 | 22.28 % | 8 | 1.40 % |
| Python | 94 | 16.49 % | 5 | 0.08 % |

A wide array of software tools is employed in BPS research and practice. Dedicated simulation software, including EnergyPlus, TRNSYS, ESP-r, IDA-ICE, and Comfie, provide robust platforms for detailed building performance analysis. Scientific software, such as Modelica, COMSOL, MATLAB and Python, equipped with specialized toolboxes or modules, is used for developing custom simulation models and performing complex data analysis (Table 1).

## 1.2   Role of computational science in BPS

Computational tools and simulations are indispensable in advancing BPS research. They allow researchers to create virtual models of buildings and systems, conduct experiments that would be impractical or impossible in the real world, and explore a vast range of scenarios to identify optimal solutions. The complexity and sophistication of simulation models in BPS have increased significantly over time. Early models were relatively simple, often focusing on single aspects of building performance. Modern BPS tools, however, integrate multiple systems and factors, including thermal dynamics, airflow, lighting, indoor air quality, acoustics, occupancy patterns, and control strategies. These models require high computational power and advanced algorithms to accurately represent the interactions between different building components, occupants, and external environmental conditions.

In this context, the imperative for reproducibility in BPS research becomes evident. Ensuring that simulation studies are reproducible is essential for validating results, advancing knowledge, and fostering trust in the scientific community. As BPS relies heavily on computational methods, the reproducibility of these methods – through sharing of code, data, computational environment and detailed methodologies – is crucial for the integrity and the progress of the field. This paper argues for the necessity of reproducible research practices in BPS and explores tools and strategies available to achieve this goal.

# 2 The need for reproducibility in building performance simulation research

## 2.1 Definition and importance of reproducibility and replicability

Reproducibility refers to obtaining consistent results by using the same data, methods, and conditions. This is also known as "computational reproducibility." Replicability means obtaining consistent results across different studies addressing the same scientific question, each with its own data. Studies are considered replicable if their results are consistent within the system's inherent uncertainty (National Academies of Sciences, 2019). Reproducible and replicable research is defined as the practice of providing all the necessary information and computational tools (including code, data and software configurations) required to replicate the results of a study (Peng, 2011). This allows other researchers to validate findings by independently recreating the analysis and confirming the results.

### 2.1.1 Reproducibility crises in science

The concept of reproducibility has gained significant attention in recent years, particularly due to several high-profile reproducibility crises across various scientific disciplines (Baker, 2016). For instance, the replication crisis in medical research revealed that many published findings could not be reproduced by independent researchers (Ioannidis, 2019; Hunter, 2017; Brownlee & Bielekova, 2021; Kafkaki, Agassi, Chesler, Crabbe, & Crusio, 2018; Jalali, DiGennaro, Guitar, Lew, & Rahmandad, 2021). These crises have underscored the need for transparency in research methodologies and the sharing of data and code.

### 2.1.2 Relevance to BPS: ensuring reliability and validity of simulation results

In the context of Building Performance Simulation (BPS), reproducibility is crucial for several reasons. First, BPS relies heavily on complex computational models and simulations to predict building behavior under various conditions. Ensuring that these simulations can be reproduced independently is essential for verifying their accuracy and reliability. Second, reproducibility fosters confidence in simulation results, making them more credible and acceptable for practical applications such as building design, policymaking, and performance optimization. Finally, reproducible research promotes knowledge sharing and collaboration within the scientific community, accelerating advancements in BPS.

## 2.2 Current state of reproducibility in BPS

### 2.2.1 Current practices in publishing code and data sharing

Major scientific publishers have data sharing policies (CHORUS, 2024). Although policy frameworks for research data sharing are proposed (Hrynaszkiewicz, Simons, Hussain, Grant, & Goudie, 2020), their number and diversity is confusing. FAIRsharing.org, which provides curated, high-quality information about metadata standards, databases and sharing policies lists more than 1700 standards, 2000 databases and 250 policies for data sharing (FAIRsharing, 2024). While data sharing requirements are becoming more common, there is limited incentive for sharing executable code and associated data. A survey of recent publications on computational and data science reveals that many studies do not provide access to the underlying code and data used in simulations (Hutson, 2018). While some journals and conferences have started to encourage or mandate the sharing of supplementary materials, the adoption of these practices remains inconsistent (Eva, Fennell, & Eve, 2021; Altman, Botgman, Crosas, & Matone, 2015; Fenner, et al., 2019). Researchers frequently publish findings based on proprietary or closed-source software, further complicating efforts to replicate their work.

*2.2.2 The gap between publication and reproducibility: unpublished code, data and computational environment*

Researchers may publish detailed descriptions of their methodologies and results, but without the actual code, data and computational environment, these descriptions are insufficient for full replication. This gap is partly due to the absence of standardized requirements for code and data sharing in many academic journals. Moreover, concerns about intellectual property, privacy, and the additional effort required to prepare code and data for public release can deter researchers from making their materials available.

To bridge this gap, the BPS community needs to embrace a culture of openness and transparency expressed, for example, in FAIR principles: Findable, Accessible, Interoperable, Reusable (Wilkinson, et al., 2016) adopted by the European Commission (Spichtinger, 2021). Journals and funding agencies should enforce stringent policies requiring the sharing of code and data. Researchers should be encouraged and supported to adopt best practices for reproducible research, such as using version control systems, providing detailed documentation, and depositing their materials in publicly accessible repositories (Peng, 2011).

# 3 Challenges to reproducibility in building performance simulation research

## 3.1 Technical challenges

*3.1.1 Complexity of simulation software and models*

Building Performance Simulation (BPS) relies on sophisticated software tools and complex models to analyze various aspects of building performance, such as energy efficiency, thermal comfort, and indoor air quality. Like in other scientific fields, the complexity of these models poses a significant challenge to reproducibility (Fitzpatrick, 2019). To enable others to replicate their work, researchers must accurately document every aspect of their simulation setup, from model parameters to computational methods. However, the intricate nature of these simulations can make comprehensive documentation labor-intensive and prone to omissions.

*3.1.2   Variability in computational environments and dependencies*

Reproducibility in BPS is further complicated by the variability in computational environments and software dependencies (Santana-Perez, et al., 2017). Many BPS tools depend on external libraries and modules, which themselves may undergo frequent updates. Managing these dependencies to ensure that simulations can be replicated is a complex task. Researchers need to provide detailed information about their computational environment, including software versions and library dependencies. The use of containerization technologies, such as Docker, can help address this challenge by encapsulating the entire computational environment, but it requires additional effort and technical expertise (Boettiger, 2015; Moreau, Wiebels, & Boettiger, 2023; Nust, et al., 2020).

*3.1.3   Insufficient documentation and standardization*

Adequate documentation is crucial for reproducible research, yet it is often insufficient or inconsistent in BPS studies. Researchers may focus primarily on publishing their findings, with less attention given to documenting the code, data, and methodologies used. This lack of detailed documentation hinders others from understanding and replicating the work. Furthermore, there is a lack of standardized practices for documentation and data sharing within the BPS community. Establishing and adhering to standardized protocols for documenting simulations and sharing data can significantly enhance reproducibility, but this requires a concerted effort from the entire research community.

## 3.2   Cultural and institutional barriers

*3.2.1   Lack of incentives and recognition for sharing code and data*

Sharing code and data requires time and effort, yet there is often little recognition or reward for these activities within the current academic system. Traditional metrics of academic success, such as citation counts and journal impact factors, do not typically account for the reproducibility of research. The academic culture of "publish or perish" places immense pressure on researchers to produce novel findings and publish them in high-impact journals. This pressure can disincentivize the additional work required to ensure reproducibility, such as thoroughly documenting code and data, or sharing these resources publicly. Researchers may prioritize the rapid publication of results over the meticulous process of making their work reproducible. To address this issue, academic institutions and funding agencies need to develop and implement metrics that value and incentivize reproducible research practices.

*3.2.2   Intellectual property and proprietary software concerns*

Concerns about intellectual property (IP) and the use of proprietary software also pose significant barriers to reproducibility in BPS. Researchers may be hesitant to share their code and data due to fears of IP theft or misuse. Encouraging the use of open-source software and developing licensing frameworks that balance IP protection with the need for reproducibility can help mitigate these concerns.

# 4 Benefits of reproducible research in building performance simulation

## 4.1 Scientific integrity and progress

Transparency fosters trust in scientific findings as researchers can verify and validate the results independently. In the context of Building Performance Simulation (BPS), where complex computational models are used to make critical decisions regarding building design and operation, ensuring transparency is paramount for stakeholders who rely on these findings to make informed choices.

Reproducibility facilitates the peer review process by enabling reviewers to assess the validity of a study based on the underlying data and methods. Reviewers can replicate the analysis and verify the results, ensuring that the research meets rigorous scientific standards.

By making code, data, and methodologies openly available to the research community, researchers can build upon existing work, avoid duplicative efforts, and collaborate more effectively. In the context of BPS, where the complexity of simulations often requires substantial computational resources and expertise, the ability to leverage shared resources can significantly enhance research productivity and innovation.

## 4.2 Educational and training value

Reproducible research serves as valuable learning resources for new researchers and students entering the field of BPS. By sharing code, data, and detailed documentation, experienced researchers provide aspiring scholars with practical examples and insights into best practices in computational research.

Reproducible research facilitates replication studies, which are essential for validating and building upon existing research findings. By providing all the necessary resources for replication, including code, data, and methodologies, researchers enable others to verify the robustness of their findings and explore new research directions. Additionally, reproducible research serves as a platform for teaching best practices in computational research.

# 5 Tools and strategies for achieving reproducibility in building performance simulation research

In computational science, where full independent replication may not be feasible, reproducibility is a minimum standard for assessing scientific claims. The spectrum of reproducibility is (Peng, 2011):

1. Publication only: This is the lowest level, where only the published paper is available, making the work essentially not reproducible.

2. Publication + code: Providing the code used for analysis along with the publication improves reproducibility, as others can inspect and verify the code.
3. Publication + code and data: Sharing both the code and data allows others to reproduce the computational workflow and results from the original data.
4. Publication + linked executable code, data and computational environment: Providing an executable environment that links the code and data together enhances reproducibility by eliminating potential issues with software dependencies or environment setup.
5. Full replication: The gold standard is full independent replication, where a study is replicated from the ground up using new data collection and independent analysis (Raphael, Sheehan, & Vora, 2020). However, this may not always be feasible for complex studies.

## 5.1 Tools for reproducibility in MATLAB

### 5.1.1 Code sharing: MATLAB Central, GitHub integration

MATLAB Central serves as a central repository for MATLAB code, allowing researchers to share their scripts, functions, and toolboxes with the community (Moler & Little, 2020). Integration with GitHub facilitates collaboration and version control, enabling researchers to track changes and manage contributions from multiple collaborators.

### 5.1.2 Generating reproducible reports: MATLAB Live Scripts

MATLAB Live Scripts offer an interactive environment for developing and documenting MATLAB code. Live Scripts combine code, visualizations, and narrative text in a single document, making it easy for researchers to explain their analyses and share their workflows with others. Researchers can also generate reproducible reports from Live Scripts, ensuring that their results can be recreated and verified by others.

### 5.1.3 Environment Management: MATLAB Runtime Container, Docker for MATLAB

MATLAB Runtime enables researchers to deploy MATLAB applications and code on systems that do not have MATLAB installed. By packaging their MATLAB code with MATLAB Runtime, researchers can distribute their simulations to collaborators or end-users without requiring them to have a MATLAB license. Additionally, Docker can be used to containerize their MATLAB code and dependencies, ensuring consistent execution across different computing environments.

## 5.2 Tools for full replication in Python

While sharing source code and data is essential for computational reproducibility, it is not sufficient on its own. Achieving true reproducibility requires multiple layers of software that run the code, including the application, its modules, and the operating system. The advantage of using open-source software, such as Python, is that it is publicly available and can be downloaded to recreate the entire computational environment. This environment can then be encapsulated in a containerization platform like Docker, ensuring that all dependencies are correctly linked and enabling precise replication of the computational setup.

### 5.2.1 Code and data: GitHub, GitLab, and generating DOIs with Zenodo

Version control platforms, like GitHub and GitLab, provide researchers with centralized repositories for storing and sharing code. By hosting their code on these platforms, researchers can facilitate collaboration, track changes, and maintain a transparent record of the development process. Additionally, platforms like Zenodo (Dillen, Grrom, Agosti, & Nielsen, 2019) allow researchers to assign a Digital Object Identifier (DOI) to their code repositories, ensuring long-term accessibility and citability.

### 5.2.2 Linked and executable code and data: mybinder.org

The website mybinder.org offers a convenient platform for running and interacting with Jupyter notebooks in the cloud (Ragan-Kelly, et al., 2018). Researchers can create reproducible workflows using Jupyter notebooks, which combine code, visualizations, and explanatory text in a single document (Granger & Perez, 2021). By sharing their notebooks via mybinder.org, researchers allow others to execute their code and explore their analyses directly on the internet using a web browser, without the need to install any software.

### 5.2.3 Environment management: conda, virtualenv, Docker

Managing computational environments is crucial for ensuring reproducibility in Python-based research. Tools like *conda* and *virtualenv* enable researchers to create isolated environments with specific dependencies, ensuring that code runs consistently across different systems (Pimentel, Murta, Braganholo, & Freire, 2019). Docker provides a more comprehensive solution by containerizing entire computational environments, including the operating system, libraries, and dependencies. By encapsulating their code and dependencies within Docker containers, researchers can guarantee that their experiments are reproducible across diverse computing environments (Boettiger, 2015; Moreau, Wiebels, & Boettiger, 2023).

### 5.2.4 Example of a reproducible and replicable paper on BPS

Let's consider as an example a "publication + linked executable code, data and computational environment" of open computational science research published in the Journal of Building Performance Simulation (Ghiaus, 2022a). The paper can be also considered as "full replication" because it allows to test the methods with data provided by the reader. The preprint is available in open access on a repository for open science (Ghiaus, 2022c). Clicking on the button 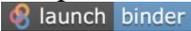 on the figures in the preprint, the reader is directed to mybinder.org website that executes remotely the code which generated the figure. The reproducibility is achieved by obtaining the same figure; the replicability is achieved by testing the algorithm on a domain of variation of input variables. The code and data used in the paper is published with a digital object identifier (DOI) on zenodo.org repository (Ghiaus, 2021). A link is provided to executable code, data and computational environment (Ghiaus, 2022b). This link allows full independent reproducibility and replication of the results of the paper in a web browser without requiring any additional software, which should be a minimum standard for judging the scientific claims (Peng, 2011).

# 6 Conclusions

The challenges and barriers to achieving reproducibility in Building Performance Simulation (BPS) are multifaceted, ranging from technical complexities to cultural and institutional norms within the academic community. Despite these challenges, the necessity of reproducible research in BPS is clear, as it ensures the reliability, validity, and credibility of scientific results, ultimately advancing our understanding of building performance and driving innovation in sustainable building practices.

Fostering a culture of openness and reproducibility is imperative for the advancement of BPS and the broader scientific community. By prioritizing reproducibility and valuing the efforts of researchers who share their code, data, and methodologies, we can create an environment conducive to rigorous and impactful research in BPS.

The Journal of Building Performance Simulation could encourage authors to submit remotely executable code and data and to request a verification of reproducibility in which the reviewers check that the code reproduces the results published in the paper. Essential elements in the article such as problem statement (e.g., boundary and initial conditions, values of associated physical and numerical parameters), validation of numerical model on standard cases, uncertainty evaluation in comparison of numerical and experimental results, numerical methods used, description of the algorithms, programming environment (versions, libraries, etc.) would be beneficial to be considered in the reviewing process. Papers that pass the reproducibility review may be highlighted in the journal, and the code and data, accessible via a persistent Digital Object Identifier (DOI), could be considered as peer-reviewed publications. By doing so, we can ensure that BPS research remains at the forefront of sustainable building design and contributes meaningfully to addressing the challenges of climate change and urbanization.

## Disclosure statement

## Data availability statement

Data supporting the findings of this viewpoint are available within the article.